\documentclass[showpacs,preprintnumbers,eqsecnum,amsmath,amssymb]{revtex4}

\usepackage{graphicx}
\usepackage{dcolumn}
\usepackage{bm}

\newcommand{\be}{\begin{equation}}
\newcommand{\ee}{\end{equation}}
\newcommand{\ba}{\begin{eqnarray}}
\newcommand{\ea}{\end{eqnarray}}

\begin{document}
\title{Can the ``brick wall" model present the same results \\
in different coordinate representations? }
 \author{Jiliang Jing} \email{jiliang.jing@ncl.ac.uk}
\affiliation{
Institute of Physics and Physics Department, \\ Hunan Normal
University,\\ Changsha, Hunan 410081, P. R. China \\
and  \\
School of Mathematics and Statistics,\\ University of Newcastle
Upon Tyne,\\ Newcastle Upon Tyne NE1 7RU, UK}

\vspace*{1.cm}
\begin{abstract}
\vspace*{0.5cm}

By using the 't Hooft's ``brick wall" model and the Pauli-Villars
regularization scheme  we calculate the statistical-mechanical
entropies arising from the quantum scalar field in different
coordinate settings, such as the Painlev\'{e} and Lemaitre
coordinates. At first glance, it seems that the entropies would be
different from that in the standard Schwarzschild coordinate since
the metrics in both the Painlev\'{e} and Lemaitre coordinates do
not possess the singularity at the event horizon as that in the
Schwarzschild-like coordinate. However, after an exact
calculation we find that, up to the subleading correction, the
statistical-mechanical entropies in these coordinates are
equivalent to that in the Schwarzschild-like coordinate. The
result is not only valid for black holes and de Sitter spaces,
but also for the case that the quantum field exerts back reaction
on the gravitational field provided that the back reaction does
not alter the symmetry of the spacetime.

\end{abstract}

\vspace*{0.2cm}

\maketitle

\section{INTRODUCTION}
\label{sec:intro} \vspace*{0.0cm}

In quantum field theory, we can use a timelike Killing vector to
define particle states. Therefore, in static spacetimes we know
that it is possible to define positive frequency modes by using
the timelike Killing vector. However, in these spacetimes there
could  arise more than one timelike Killing vector which make the
vacuum states inequivalent. This means that the concept of
particles is not generally covariant in curve spacetime.

Bekenstein and Hawking\cite{Bekenstein72} \cite{Hawking75} found
that, by comparing black hole physics with thermodynamics and
from the discovery of black hole evaporation, black hole entropy
is proportional to the area of the event horizon. The discovery is
one of the most profound in modern physics. However, the issue of
the exact statistical origin of the black hole entropy has
remained a challenging one. Recently, much effort has been
concentrated on the problem \cite{Solodukhin95} -\cite{Mann96}.
The ``brick wall" model (BWM) proposed by 't Hooft \cite{Hooft85}
is an extensively used way to calculate the entropy in a variety
of black holes, black branes, de Sitter spaces, and anti-de
Sitter spaces \cite{Hooft85} -\cite{Mann96}. In this model the
Bekenstein-Hawking entropy of the black hole is identified with
the statistical-mechanical entropy arising from a thermal bath of
quantum fields propagating outside the event horizon.

The concept of particles in quantum field theory is not generally
covariant and depends on the coordinate representations. This
leads to an interest question: can we get the same results for
statistical-mechanical entropy of black holes in different
coordinate representations, such as the Painlev\'e and Lemaitre
coordinates,  by employing the BWM by making use of the wave modes
in this model? At first sight, we might anticipate that the
results are different since the wave modes obtained by using
semiclassical techniques are the exact modes of the quantum
system in the asymptotic region. Thus, if the asymptotic
structures of the spacetime are different for any two
coordinates, then the semiclassical wave modes associated with
different coordinates will be different. The aim of this paper is
to study this question carefully by applying the BWM to two
different coordinate representations of the general standard
static black hole and studying the statistical-mechanical
entropy. The two coordinate representations which we use are the
stationary Painlev\'e coordinate and the time dependent Lemaitre
coordinate. In both Painlev\'e and Lemaitre coordinates, the
metrics have no coordinate singularity which are different from
the standard Schwarzschild-like coordinate. However, they both
acquire singularity at the event horizon in the action function.
Therefore, there could be particle production in these
coordinates and hence we can use the knowledge of the wave modes
of the quantum field in these coordinate settings to calculate the
statistical-mechanical entropies.

In order to compare the statistical-mechanical entropies obtained
in this paper with the result for the standard Schwarzschild-like
coordinate, we first introduce the expression of the entropy for
the Schwarzschild-like coordinate in the following. In the BWM, in
order to eliminate divergence which appears due to the infinite
growth of the density of states close to the horizon, 't Hooft
introduces a brick wall cutoff: a fixed boundary $\Sigma_h$ near
the event horizon within the quantum field does not propagate and
the Dirichlet boundary condition was imposed on the boundary, i.
e., wave function $\phi=0$ for $r=r({\Sigma_h})$. However,
Demers, Lafrance, and Myers \cite{Demers95} found, in the
Pauli-Villars regulated theory, that 't Hooft's brick wall can be
removed  by introducing five regulator fields: $\phi_1$ and
$\phi_2$, which are two anticommuting fields with mass
$m_1=m_2=\sqrt{\mu^2+m^2}$ (where $\mu$ represents the UV
cutoff); $\phi_3$ and $\phi_4$, which are two commuting fields
with mass $m_3=m_4=\sqrt{3 \mu^2+m^2}$; and $\phi_5$, which is an
anticommuting field with mass $m_5=\sqrt{4\mu^2+m^2}$. Together
with the original scalar field $\phi=\phi_0$ with mass $m=m_0$
these fields satisfy the two constraints $\sum^5_{i=0}\Delta_i=0$
and $\sum^5_{i=0}\Delta_i m_i^2=0$, where $\Delta_i=+1$ for the
commuting fields, and $\Delta_i=-1$ for the anticommuting fields.
By using the BWM and Pauli-Villars regulators, Demers, Lafrance
and Myers \cite{Demers95}, and Solodukhin \cite{Solodukhin97}
found  that the statistical-mechanical entropy arising from the
minimally coupled quantum scalar field in a general nonextreme
static black hole
\begin{eqnarray}
ds^2=-g(r)dt_s^2+\frac{1}{g(r)}dr^2+R^2(r)(d\theta ^2+\sin^2\theta
d\varphi ^2) \label{Sch}
\end{eqnarray}
(where $g(r)$ is an arbitrary function of $r$. The event horizon
is determined by $g(r)=0$. And $(dg(r)/dr) |_{r_+}\neq 0$ for the
nonextreme black holes) can be expressed as
\begin{eqnarray}
    S&=&\frac{A_\Sigma}{48\pi}\sum_{i=0}^5\Delta_i m_i^2 \ln
    m_i^2\nonumber \\
&-&\frac{A_{\Sigma}}{288\pi}\left[{\mathcal{R}}-\frac{1}{5}
\left(\frac{\partial^2 g(r)}{\partial^2 r}-\frac{1}{R^2(r)}
\frac{\partial g(r)}{\partial r}\frac{\partial R^2(r)}
{ \partial r}\right)\right]_{r_+}\sum_{i=0}^5\Delta_i
\ln m_i^2,\label{Schen}
\end{eqnarray}
where $A_\Sigma=\int d\theta d\varphi [\sqrt{g_{\theta\theta}
g_{\varphi\varphi}}]_{r_+}$ is the area of the event horizon,
${\mathcal{R}}$ is a scalar curvature of the spacetime.  The
statistical-mechanical entropy (\ref{Schen}) obtained by this
approach consists of two parts: the first part, after taking
renormalization of the gravitational constant as
$\frac{1}{G_R}=\frac{1}{G_B}+\frac{1}{12\pi}\sum_{i=0}^5\Delta_i
m_i^2 \ln m_i^2$,  gives Bekenstein-Hawking entropy,  and the
second part can be considered as a quantum correction to the
entropy of the black hole due to the quantum scalar field.

The paper is organized as follows. In sec. II, the Painlev\'e
spacetime is introduced and the statistical-mechanical entropy
arises from the quantum scalar field in the Painlev\'e coordinate
that is studied. In sec. III,  the statistical-mechanical entropy
due to the quantum scalar field in the Painlev\'e coordinate is
investigated. The summary and discussions are presented in sec.
IV.

\section{Statistical-mechanical entropy in the Painlev\'{e} coordinate }
 \vspace*{0.0cm}

We now investigate statistical-mechanical entropy that arises from
the quantum scalar field in the Painlev\'e coordinate system. The
time coordinate transformation from the standard
Schwarzschild-like coordinate (\ref{Sch}) to the Painlev\'e
coordinate is
\begin{eqnarray}
    t=t_s+\int \frac{\sqrt{1-g(r)}}{g(r)} dr.
\end{eqnarray}
The radial and angular coordinates remain unchanged.  With this
choice, the line element (\ref{Sch}) becomes
\begin{eqnarray}
ds^2=-g(r)dt^2+2\sqrt{1-g(r)}dt dr+dr^2+R^2(r)(d\theta ^2+
\sin^2\theta d\varphi ^2), \label{Pain}
\end{eqnarray}
which is the Painlev\'e coordinate representation. The coordinate
has distinguishing features: (a) The  spacetime is stationary but
not static; (b) the constant-time surfaces is flat if
$R^2(r)=r^2$; And (c) there is now no singularity at $g(r)=0$.
That is to say, the coordinate complies with the perspective of a
free-falling observer, who is expected to experience nothing out
of the ordinary upon passing through the event horizon. However,
the event horizon manifests itself as a singularity in the
expression for the semiclassical action. It is easily to prove
that the inverse Hawking temperature
\begin{eqnarray}
 \beta_H=2\pi\left.\frac{1+\sqrt{1-g(r)}}{\frac{ d g(r)}{d r}}\right|_{r_+}=4\pi/\left.\frac{ d g(r)}{d r}\right|_{r_+},
\end{eqnarray}
 is recovered in the Painlev\'e coordinate by using the complex path technique \cite{Shan1} \cite{Shan2}.

We now try to find an expression of the statistical-mechanical
entropy due to the quantum scalar field in thermal equilibrium at
temperature $1/\beta$ in the Painlev\'{e} coordinate by suing the
BWM. Using the WKB approximation with
\begin{equation}
\phi =exp[-iEt+iW(r,\theta, \varphi)], \label{phi}
\end{equation}
and substituting  the metric (\ref{Pain}) into  the Klein-Gordon
equation of the scalar field $\phi$ with mass  $m $ and
nonminimal $\xi {\mathcal{R}} \phi $ ({ $\mathcal{R}$ } is the
scalar curvature of the spacetime) coupling
\begin{equation}
\frac{1}{\sqrt{-\tilde{g}}}\partial _\mu(\sqrt{-\tilde{g}}
g^{\mu\nu}\partial _\nu \phi)-(m^2+\xi {\mathcal{R}})=0,
\label{kg}
\end{equation}
we find
\begin{equation}
p_r^{\pm}=\frac{1}{g(r)} \left[\sqrt{1-g(r)} E \pm \sqrt{g(r)}
\sqrt{\frac{E^2}{g(r)}
-\left(\frac{p^2_\theta}{R^2(r)}+\frac{p^2_{\varphi}}{R^2(r)
\sin^2\theta}+M^2(r)\right)} \right],\label{W}
\end{equation}
 where $p_r\equiv \partial _r W(r, \theta, \varphi)$,  $p_\theta \equiv
 \partial_\theta W(r, \theta, \varphi)$, and  $p_\varphi \equiv
 \partial_\varphi W(r, \theta, \varphi)$ are the momentum
 of the particles moving in $r$, $\theta$, and $\varphi$,
 respectively. The sign ambiguity of the square root is related
 to the ``out-going"  ($p_r^+$) or ``in-going"  ($p_r^-$)
 particle, respectively.  If the scalar curvature ${\mathcal{R}}$
 takes a nonzero value at the horizon then this region can be
 approximated by some sort of constant curvature space. However,
 the exact result for such a black hole showed that the mass
 parameter in the solution enters only in the combination
 $(m^2-{\mathcal{R}}/6)$ \cite{Solodukhin97} \cite{Birrell82},
 and then $ M^2(r)=m^2-(\frac{1}{6}-\xi){\mathcal{R}}$ in the
 equation (\ref{W}). In this paper our discussion is restricted
 to study minimally coupled $(\xi=0)$ scalar fields since the main
aim of this paper is to see whether the brick wall model can
present the same result in different coordinates.

 The partition function is given by
\begin{eqnarray}
    z=\sum_{n_q}exp[-\beta(E_q) n_q],
\end{eqnarray}
where $q$ denotes a quantum state of the field with energy $E_q$.
The free energy is
\begin{eqnarray}
    F&=&\frac{1}{\beta}\int d p_{\theta} \int d p_{\varphi} \int
d n(E, p_{\theta}, p_{\varphi}) \ln \{1-exp[-\beta E]\}\nonumber \\
    &=&-\int d p_{\theta} \int d p_{\varphi} \int \frac{n(E,
     p_{\theta}, p_{\varphi})}{e^{\beta E}-1} d E\nonumber \\
    &=&- \int \frac{n(E)}{e^{\beta E}-1} d E
    \label{free}
\end{eqnarray}
where $n(E)\equiv \int dp_{\theta}\int d p_{\varphi} n(E,
p_{\theta},  p_{\varphi})$ presents the total number of the modes
with energy less than $E$. In phase space the total number of
modes with $E$ is given by
\begin{eqnarray}
     n(E)&=&\frac{1}{\pi}\int d\theta \int _{r_++h}^{L} dr\int
      d p_{\theta}d p_{\varphi} \frac{p_r^+-p_r^-}{2} \nonumber \\
     &=&\frac{1}{\pi}\int d\theta \int _{r_++h}^{L} dr \int
     d p_{\theta}d p_{\varphi}  \frac{1}{\sqrt{g(r)}}
     \sqrt{\frac{E^2}{g(r)}-\left(\frac{p^2_\theta}{R^2(r)}+
     \frac{p^2_{\varphi}}{R^2(r)\sin^2\theta}+M^2(r) \right)} .
     \nonumber\\ \label{nn1}
\end{eqnarray}
The integral is taken only over those values for which the square
root exists. In Eq. (\ref{nn1}) we utilize the average of the
radial momentum (the minus before the $p_r^-$ is caused by a
different direction). In this way, the total number of modes is
related to all kinds of particles. We checked that this
definition can also be used for all previous corresponding works.
Carrying out the integrations of the $p_{\theta}$, $p_{\varphi}$,
and $r$, we get
\begin{eqnarray}
n(E)=&-&\frac{1}{2\pi}\int d\theta \left\{ \sqrt{g_{\theta\theta}
 g_{\varphi\varphi}} \left[ \frac{2}{3}\left(\frac{\beta_H E}{4\pi}
 \right)^3 C(r)+ M^2(r)\left(\frac{ \beta_H E}{4 \pi} \right) \right]
 \ln \frac{E^2}{E^2_{min}} \right\}_{r_+}\nonumber\\
&-& \frac{1}{3 \pi}\frac{ \beta_H}{4 \pi} \int d\theta \left[
\sqrt{g_{\theta\theta} g_{\varphi\varphi}} M^2(r)\left(E-\frac{E^3}
{E^2_{min}}\right) \right]_{r_+}, \label{ne}
\end{eqnarray}
where
\begin{eqnarray}
    C(r)&=&\frac{\partial^2 g(r)}{\partial^2 r}-\frac{1}{R^2(r)}
    \frac{\partial g(r)}{\partial r}\frac{\partial R^2(r)}
    { \partial r}, \nonumber \\
    E^2_{min}&=&[M^2(r)g(r)]_{\Sigma_h}.
\end{eqnarray}
We now use the Pauli-Villars regularization scheme introduced in
the preceding section. Since each of the scalar fields makes a
contribution to the free energy, the total free energy can be
expressed as
\begin{eqnarray}
    \beta \bar{F}=\sum_{i=0}^5\beta \Delta_i F_i. \label{Sf}
\end{eqnarray}
Substituting  Eqs. (\ref{free}) and (\ref{ne}) into Eq. (\ref{Sf})
and then taking the integration over $E$ we have
\begin{eqnarray}
\bar{F}&=&-\frac{1}{48\pi }\frac{\beta_H}{\beta ^2}\int
d\theta d\varphi \left\{\sqrt{g_{\theta \theta}g_{\varphi
\varphi}}\right\}_{r_+} \sum^5_{i=0}\triangle _i
M_i^2(r_H)lnM_i^2(r_H) \nonumber \\ & &
-\frac{1}{2880\pi}\frac{\beta_H^3}{\beta^4} \int
d\theta d \varphi \left\{\sqrt{g_{\theta \theta}g_{\varphi \varphi}}
\left[\frac{\partial
^2g(r)}{\partial r^2}-\frac{1}{R^2(r)}\frac{\partial g(r)}{\partial r}
\frac{\partial R^2(r)}{\partial r}\right]\right\}_{r_+}
 \sum^5_{i=0}\triangle_i \ln M_i^2(r_H).\nonumber \\  \label{f-0}
\end{eqnarray}
Using  the assumption that the scalar curvature ${\mathcal{R}}$ at
the horizon is much smaller than each $m_i$ and inserting free
energy into the relation
\begin{eqnarray}
    S=\beta^2 \frac{\partial F}{\partial \beta},
\end{eqnarray}
we obtain the expression of the statistical-mechanical entropy due
to a minimally coupled scalar field in the Painlev\'e coordinates
\begin{eqnarray}
    S&=&\frac{A_\Sigma}{48\pi}\sum_{i=0}^5\Delta_i m_i^2 \ln m_i^2
    \nonumber \\
&-&\frac{A_{\Sigma}}{288\pi}\left[{\mathcal{R}}-\frac{1}{5}
\left(\frac{\partial^2 g(r)}{\partial^2 r}-\frac{1}{R^2(r)}
\frac{\partial g(r)}{\partial r}\frac{\partial R^2(r)}{ \partial r}
\right)\right]_{r_+}\sum_{i=0}^5\Delta_i \ln m_i^2,\label{Painen}
\end{eqnarray}
where $A_{\Sigma}= \int d\varphi  d\theta\left\{\sqrt{g_{\theta
\theta}g_{\varphi \varphi}}\right\}_{r_+}=4\pi R^2(r_+)$ is the
area of the event horizon.

By the equivalence principle and the standard quantum field
theory in flat space, to construct a vacuum state for the
massless scalar field in the Painlev\'e spacetime we should leave
all the positive frequency modes empty. Kraus \cite{Kraus}
pointed out that for the metric (\ref{Pain}) it is convenient to
work along a curve
\begin{eqnarray}
    dr+\sqrt{1-g(r)}dt=0, \label{condition}
\end{eqnarray}
then the condition is simply a positive frequency with respect to
$t$ near this curve.  It is easy to prove that the modes used to
calculate the entropy are essentially the same as that in the
Schwarzschild-like coordinates. Therefore, it is reasonable that
the result (\ref{Painen}) is exactly equal to entropy
(\ref{Schen}).

\section{Statistical-mechanical entropy in The Lemaitre coordinate }
 \vspace*{0.0cm}

In this section we study statistical-mechanical entropy due to
the  quantum scalar field in the Lemaitre coordinates. The
coordinates that transform from the Painlev\'e coordinates
(\ref{Pain}) to the Lemaitre coordinates are given by
\begin{eqnarray}
 \tilde{r}&=&t+\int \frac{d r}{\sqrt{1-g(r)}}, \nonumber \\
    U&=&\tilde{r}-t, \nonumber \\
    V&=&\tilde{r}+t, \label{chan}
\end{eqnarray}
where $t$ is the Painlev\'e time. The angular coordinates
$\theta$ and $\varphi$  remain the same. The line element
(\ref{Pain}), in the new coordinates, is described by
\begin{eqnarray}
    ds^2=\frac{(f(U)-1)}{4}(d V^2+d U^2)+\frac{(f(U)+1)}{2}d V
    d U +y(U)(d \theta ^2+\sin^2 \theta d \varphi^2), \label{Lema}
\end{eqnarray}
where
\begin{eqnarray}
 f(U)& \equiv & 1-g(r), \nonumber \\
 y(U)& \equiv & R^2(r). \label{gtof}
\end{eqnarray}
The line element (\ref{Lema}) is the Lemaitre coordinate
representation of the spacetime (\ref{Sch}). The metric in the
Lemaitre coordinate is no singularity at $g(r)=0$ just as in the
Painlev\'{e} coordinates. However, the horizon also manifests
itself as a singularity in the expression for the semiclassical
action. We can also show that the inverse Hawking temperature
\begin{eqnarray}
    \beta_H=-\pi\left.\frac{(1+\sqrt{f})^2}{\frac{\partial f}{\partial
U}}\right|_{U_0}=4\pi/\left.\frac{ d g(r)}{d
r}\right|_{r_+}, \label{temp}
\end{eqnarray}
is recovered in the Lemaitre
coordinate by employing the complex path technique  \cite{Shan1}
\cite{Shan2}. In Eq. (\ref{temp})  $U_{0}$ represents the root of
the equation $(1-f)=g=0$.

We can use the WKB approximation with
\begin{equation}
\phi =exp[-iEV+iW(U,\theta, \varphi)]. \label{phi1}
\end{equation}
The reason for using the modes with positive frequency with
respect to the coordinate $V$ is that another coordinate
$U=\tilde{r}-t= \int\frac{d r}{\sqrt{1-g(r)}}$ is related to the
space coordinate $r$ of the original coordinates only.

Substituting  Eq. (\ref{phi1}) and metric (\ref{Lema}) into  the
Klein-Gordon equation of the scalar field with mass $m$, Eq.
(\ref{kg}), we have
\begin{equation}
p_U^{\pm}=\frac{f}{1-f}
\left[\frac{1+f}{f} E \pm \sqrt{\frac{1-f}{f}} \sqrt{\frac{4E^2}{1-f}
-\left(\frac{p^2_\theta}{y}+\frac{p^2_{\varphi}}{y\sin^2\theta}+M^2(U)
\right)} \right],\label{W1}
\end{equation}
 where $p_U\equiv \partial _U W(U, \theta, \varphi)$,  $p_\theta \equiv
 \partial_\theta W(U, \theta, \varphi)$ and  $p_\varphi \equiv
 \partial_\varphi W(U, \theta, \varphi)$ are the momentum of the particle
 moving in $U$, $\theta$ and $\varphi$, respectively, and
 $M^2(U)=m^2-\frac{1}{6}{\mathcal{R}}$. Therefore, in phase space
 we obtain the number of modes
\begin{eqnarray}
     n(E)&=&\frac{1}{\pi}\int d\theta d\varphi \int _{U_0+\tilde{h}}
     ^{\tilde{L}} dU \int d p_\theta d p_\varphi \frac{p_U^+-p_U^-}{2}
      \nonumber \\
 &=&\frac{2}{\pi}\int d\theta d\varphi \int _{U_0+\tilde{h}}^{\tilde{L}}
  dU \int d p_\theta d p_\varphi  \sqrt{\frac{f}{1-f}} \sqrt{\frac{E^2}
  {1-f}-\left(\frac{p^2_\theta}{4 y}+\frac{p^2_{\varphi}}{4 y\sin^2\theta}
  +\frac{M^2(U)}{4}\right)}\nonumber  \\ \label{nn2}
\end{eqnarray}
where we make use of the average of the $U$-direction momentum
(the minus before the $p_U^-$ is caused by a different direction).
The integral in the second line is taken only over those values
for which the square root exists. Carrying out the integrations of
the $p_{\theta}$, $p_{\varphi}$, and $U$, we get
\begin{eqnarray}
n(E)=&-&\frac{1}{2\pi}\int d\theta \left\{ \sqrt{g_{\theta\theta}
g_{\varphi\varphi}} \left[ \frac{2}{3}\left(\frac{\beta_H E}{4\pi}
\right)^3 \tilde{C}(U)+ M^2(U)\left(\frac{ \beta_H E}{4 \pi} \right)
\right] \ln \frac{E^2}{E^2_{min}} \right\}_{U_0}\nonumber\\
&-& \frac{1}{3 \pi}\frac{ \beta_H}{4 \pi} \int d\theta \left[
\sqrt{g_{\theta\theta} g_{\varphi\varphi}} M^2(U)\left(E-
\frac{E^3}{E^2_{min}}\right) \right]_{U_0}, \label{nel}
\end{eqnarray}
where
\begin{eqnarray}
    \tilde{C}(U)&=&\frac{1}{ f}\frac{\partial^2 f}{\partial^2 U}-
    \frac{1}{2 f^2}\left(\frac{\partial f}{\partial U}\right)^2-
    \frac{1}{ f y}\frac{\partial f}{\partial U}\frac{\partial y}
    { \partial U}, \nonumber \\
    E^2_{min}&=&[M^2(U_0)(1-f)]_{\Sigma_h}.
\end{eqnarray}
We now introduce the Pauli-Villars regularization scheme as
before. Substituting  Eqs. (\ref{free}) and (\ref{nel}) into Eq.
(\ref{Sf}) and then taking the integration over $E$ we have
\begin{eqnarray}
\bar{F}&=&-\frac{1}{48\pi }\frac{\beta_H}{\beta ^2}\int
d\theta d\varphi \left\{\sqrt{g_{\theta \theta}g_{\varphi
\varphi}}\right\}_{U_0} \sum^5_{i=0}\triangle _i
M_i^2(U_0)lnM_i^2(U_0) \nonumber \\ & &
-\frac{1}{2880\pi}\frac{\beta_H^3}{\beta^4} \int
d\theta d \varphi \left\{\sqrt{g_{\theta \theta}g_{\varphi \varphi}}
 \left[
\frac{1}{ f}\frac{\partial^2 f}{\partial^2 U}-\frac{1}{2 f^2}\left
(\frac{\partial f}{\partial U}\right)^2
  \right. \right. \nonumber \\ & & - \left. \left. \frac{1}{ f y}
  \frac{\partial f}{\partial U}\frac{\partial y}{ \partial U}
\right]\right\}_{U_0}
 \sum^5_{i=0}\triangle_i \ln M_i^2(U_0).\nonumber \\  \label{f-1}
\end{eqnarray}
Using  the assumption that the scalar curvature ${\mathcal{R}}$
at the horizon is much smaller than each $m_i$ and inserting free
energy into the relation $    S=\beta^2 \frac{\partial F}{\partial
\beta} $, we obtain the expression of the statistical-mechanical
entropy in the Lemaitre coordinate
\begin{eqnarray}
    S&=&\frac{A_\Sigma}{48\pi}\sum_{i=0}^5\Delta_i m_i^2 \ln m_i^2
    \nonumber \\
&-&\frac{A_{\Sigma}}{288\pi}\left\{{\mathcal{R}}-\frac{1}{5}\left[
\frac{1}{ f}\frac{\partial^2 f}{\partial^2 U}-\frac{1}{2 f^2}
\left(\frac{\partial f}{\partial U}\right)^2-\frac{1}{ f y}
\frac{\partial f}{\partial U}\frac{\partial y}{ \partial U}
\right]\right\}_{r_+}\sum_{i=0}^5\Delta_i \ln m_i^2,\label{Lemaen}
\end{eqnarray}
where $A_\Sigma=4\pi y|_{U_0}=4\pi R^2(r_+)$ is the horizon
area.

By using Eq. (\ref{chan}), it is easy to prove
\begin{eqnarray}
&&\left[\frac{1}{ f}\frac{\partial^2 f}{\partial^2 U}-\frac{1}{2 f^2}
\left(\frac{\partial f}{\partial U}\right)^2-\frac{1}{ f y}
\frac{\partial f}{\partial U}\frac{\partial y}{ \partial U}
\right]_{U_0}\nonumber\\   &=&\left[\frac{\partial^2 g(r)}
{\partial^2 r}-\frac{1}{R^2(r)}\frac{\partial g(r)}{\partial r}
\frac{\partial R^2(r)}{ \partial r}\right]_{r_+}.
\end{eqnarray}
This shows that the result (\ref{Lemaen}) for the Lemaitre coordinate
is equal to entropy (\ref{Painen}) for the Painlev\'{e} coordinate,
and the entropy (\ref{Schen}) for the standard Schwarzschild coordinate.
It is
well-known that the wave modes obtained by using semiclassical
techniques, in general, are the exact modes of the quantum system
in the asymptotic regions. Thus, if the asymptotic structure of
the spacetime is the same for the two coordinates, then the
semiclassical wave modes associated with these two coordinate
systems will be the same.  From Eq. (\ref{chan}) we know that the
differential relationship between the Lemaitre time $V$ and the
Painlev\'e time $t$ can be expressed as
\begin{eqnarray}
 d V=d t+d \tilde{r}=2 d t + \frac{dr}{\sqrt{1-g(r)}}.  \label{dV}
\end{eqnarray}
Now let us also work along the curve $dr+\sqrt{1-g(r)}dt=0$,
equation (\ref{dV}) then becomes
\begin{eqnarray}
    dV=dt.
\end{eqnarray}
It is shown that the two definitions of positive frequency --
with respect to $V$ in the Lemaitre spacetime and with respect
to $t$ in the Painlev\'{e} spacetime -- do coincide. Therefore,
it should not be surprised at the entropies driven from the modes
in the Lemaitre and  Painlev\'{e} coordinates are the same.

\section{Summary and discussions}

We have investigated the statistical-mechanical entropies arising
from the quantum scalar field in the Painlev\'{e} and Lemaitre
coordinates by using the 't Hooft brick wall model and the
Pauli-Villars regularization scheme. At first glance, we might
have anticipated that the results are different from that of the
standard Schwarzschild coordinate due to two reasons: a) both the
Painlev\'{e} and Lemaitre spacetimes possess a distinguishing
property: the metrics do not possess singularity at event
horizon; b) it is not obvious that the time $V$ in the Lemaitre
spacetime tends to the time $t$ in the  Painlev\'e spacetime.
Nevertheless, for either the Painlev\'{e} or Lemaitre coordinate,
the event horizon manifests itself as a singularity in the action
function and then there could be particles production. Hence we
can use the knowledge of the wave modes of the quantum field to
calculate the statistical-mechanical entropies. By comparing our
results (\ref{Painen}) and (\ref{Lemaen}), which are worked out
exactly, with the well-known  result (\ref{Schen}) we find that,
up to a subleading correction, the statistical-mechanical
entropies arising from the quantum scalar field in both the
Painlev\'e and Lemaitre coordinates are equivalent to that in the
standard Schwarzschild-like coordinate. When we construct a
vacuum state for the massless scalar field in the Painlev\'e
spacetime we take the condition $dr+\sqrt{1-g(r)}dt=0$, and then
we find that the modes used to calculate the entropies in the both
Painlev\'{e} and Lemaitre coordinates are essentially the same as
that in the Schwarzschild-like coordinates since both $V$ and $t$
tend to the Schwarzschild time $t_s$ as $r$ goes to infinity
under this condition. Therefore, it should not be surprise that
the entropies driven from the modes in the Lemaitre, Painlev\'{e},
and Schwarzschild coordinates are the same.

We should note that all the results are obtained based alone on
the most general metric (\ref{Sch}) and the conditions
$g(r)|_{r_+}=0$ and $\frac{d g(r)}{d r}|_{r_+}\neq 0$ (nonextreme
black hole). Therefore, the results are valid not only for the
spacetimes that we have known, such as the Schwarzschild, the
Reissner-Nordstr\"om, the Garfinkle-Horowitz-Strominger dilaton
\cite{Garfinkle}, the Gibbons-Maeda dilaton \cite{Gibbons}, the
Garfinkle-Horne dilaton \cite{Garfinkle1} black holes, and the
Schwarzschild de Sitter and the Reissner-Nordstr\"om de-Sitter
spaces, etc., but also for the case that the quantum field exerts
back reaction to the gravitational field provided that the back
reaction does not alter the symmetry of the spacetime.

\begin{acknowledgments}
I would like to thank Professor Ian G. Moss for helpful comments.
This work was supported by the National Natural Science Foundation
of China under Grant No. 10275024;  and the FANEDD under Grant
No. 2003052.
\end{acknowledgments}


\begin{thebibliography}{99}

\bibitem{Bekenstein72}J. D. Bekenstein, \ Lett.
Nuovo \ Cimento  {\bf 4}, 737 (1972); Phys. \ Rev. \ D {\bf 7},
2333 (1973).

\bibitem{Hawking75}S. W. Hawking,
Nature (London) {\bf 248}, 30 (1974); Commun. \ Math. \ Phys. {\bf
43}, 199 (1975).


\bibitem{Solodukhin95}S. N. Solodukhin,
Phys.\ Rev.\ D {\bf 52}, 7046 (1995);  ibid., {\bf D 54}, 3900
(1996).

\bibitem{Frolov98}V. P. Frolov and D. V. Fursaev,
Class. \ Quantum \ Grav.  {\bf 15}, (1998) 2041.

\bibitem{Jing99a}Jiliang Jing and Mu-Lin Yan,
 \ Phys. \ Rev. D  {\bf  62},  104013
(2000); {\sl ibid.,}  {\bf 63}, 024003 (2001).


\bibitem{Barvinsky95}A. O. Barvinsky, V. P. Frolov, and A. I. Zelnikov,
Phys.\ Rev.\ D {\bf 51}, 1741 (1995).

\bibitem{Belgiorno96}F. Belgiorno and S. Liberati,
Phys.\ Rev.\ D {\bf 53}, 3172 (1996).

\bibitem{Cognola980}G. Cognola and P. Lecca,
Phys.\ Rev.\ D {\bf 57}, 1108 (1998).

\bibitem{Frolov}V. P. Frolov, D. V. Fursaev, and A. I. Zelnikov,
Phys.\ Lett.\ B {\bf 382}, 220 (1996).


\bibitem{Cognola98}G. Cognola,
Phys. \ Rev.  D {\bf 57}, 6292 (1998).


\bibitem{Hooft85}G. 't. Hooft,
Nucl. \ Phys.\ B {\bf 256}, 727 (1985).

\bibitem{Jing98}Jiliang Jing,
Int. J. Theor. Phys.  {\bf 37}, 1441 (1998).


\bibitem{Jing99}Jiliang Jing and Mu-Lin Yan,
 \ Phys. \ Rev. D {\bf 60}, 084015 (1999);
{\sl ibid.,}  {\bf 61}, 044016 (2000).


\bibitem{Ghosh94}A. Ghosh, and  P. Mitra,
Phys.\ Rev.\ Lett. {\bf 73}, 2521 (1994).

\bibitem{Lee96}M.H. Lee, and J. K. Kim,
Phys.\ Lett.\ A {\bf 212}, 323 (1996); Phys.\ Rev.\ D {\bf 54},
3904 (1996).


\bibitem{Jing2000I}Jiliang Jing, and Mu-Lin Yan,
 Int. J. Theor. Phys. 39, 1687 (2000).

\bibitem{Jing2001}Jiliang Jing, and Mu-Lin Yan,
Phys. \ Rev.  D {\bf 63}, 084028 (2001).

\bibitem{Jing2001c}Jiliang Jing, and Mu-Lin Yan,
Phys. \ Rev. D {\bf 64} 064015 (2001).


\bibitem{Ho96}J. Ho, W. T. Kim, and Y. J. Park,
Class. Quantum Grav. {\bf 14}, 2617 (1997); J. Ho and G. Kang,
Phys. \ Lett. \ B  {\bf 445}, 27 (1998).


\bibitem{Bai03} Hua Bai, Mu-Lin Yan,
 Remarks on 't Hooft's Brick Wall Model, gr-qc/0303006.

\bibitem{Gupta03} Kumar S. Gupta, Siddhartha Sen,
 Hidden Degeneracy in the Brick Wall Model of Black Holes, hep-th/0302183.

\bibitem{wu03} S. Q. Wu, M. L. Yan,
 Entropy of Kerr-de Sitter black hole due to arbitrary spin fields,  gr-qc/0303076.

\bibitem{Cha02} Kyung-Seok Cha, Bum-Hoon Lee, Chanyong Park,
dS/CFT correspondence from the Brick Wall method, hep-th/0207194.

\bibitem{Li02} Li Xiang, Phys. \ Lett. B{\bf 540}  9 (2002).

\bibitem{Medved02} A.J.M. Medved,
Class. \ Quant. \ Grav. 19, 405 (2002).

\bibitem{Kim01} Won Tae Kim, John J. Oh, Young-Jai Park, Phys. \ Lett. B{\bf 512}  131 (2001).

\bibitem{Solodkhin95} S. N. Solodukhin, phys. rev. D {\bf 51}, 609 (1995).

\bibitem{Winstanley} Elizabeth Winstanley,
 Phys.Rev. D{\bf 63} 084013 (2001).

\bibitem{Kim99} Won Tae Kim, Phys.\ Rev. D{\bf 59} 047503 (1999).


\bibitem{Mann96}R. B. Mann and S. N. Solodukhin,
Phys. \ Rev.  D{\bf 54}, 3932 (1996).

\bibitem{Demers95} J. G. Demers, R. Lafrance, and R. C. Myers, Phys. \ Rev. D {\bf 52}, 2245 (1995).

\bibitem{Solodukhin97} S. N. Solodukhin, Phys. \ Rev. D {\bf 56}, 4968 (1997).

\bibitem{Birrell82} R. B. Birrell and P. C. W. Devies, Quantum fields in curved space, (Cambridge University Press, Cambridge, England, 1982).

\bibitem{Shan1}S. Shankaranarayanan, K. Srinivasan, and T. Padmanabhan,
Mod. Phys. Lett. A {\bf 16}, 571 (2001).

\bibitem{Shan2}S. Shankaranarayanan, T. Padmanabhan, and K. Srinivasan,
Class. Quantum Grav. {\bf 19}, 2671  (2002).

\bibitem{Kraus} P. Kraus, Some applications of a simple stationary line element for the Schwarzschild geometry. gr-qc/9406042.

\bibitem{Garfinkle} D. Garfinkle, G. T. Horowitz, and A. Strominger, Phys. Rev. D, {\bf 43} 3140 (1991).

\bibitem{Gibbons} G. W. Gibbons and K. Maeda, Nucl. Phys. B, {\bf 298} 741 (1988).

\bibitem{Garfinkle1}J. H. Horne and G. T. Horowitz, Phys. Rev. D, {\bf 46} 1340 (1992).


\end{thebibliography}
\end{document}